\documentclass[prl,twocolumn,superscriptaddress,amsmath,amssymb]{revtex4-1}
\usepackage{graphicx}
\usepackage{units}
\usepackage{color}
\usepackage[dvipsnames]{xcolor}
\usepackage{eqnarray}
\usepackage{tabularx}
\usepackage{bbm}
\usepackage{changes}
\usepackage{soul}

\begin{document}

    \title{Flat optical conductivity in ZrSiS due to two-dimensional Dirac bands}
    \author{M. B. Schilling}
    \affiliation{1.~Physikalisches Institut, Universit\"at Stuttgart, Pfaffenwaldring 57, 70569 Stuttgart, Germany}
    \author{L. M. Schoop}
    \author{B. V. Lotsch}
    \affiliation{Max Planck Institute for Solid State Research, Heisenbergstr. 1, 70569 Stuttgart, Germany}
    \author{M. Dressel}
    \author{A. V. Pronin}
    \affiliation{1.~Physikalisches Institut, Universit\"at Stuttgart, Pfaffenwaldring 57, 70569 Stuttgart, Germany}
    \date{July 30, 2017}

\begin{abstract}

ZrSiS exhibits a frequency-independent interband conductivity
$\sigma(\omega) = \rm{const}(\omega) \equiv \sigma_{\rm{flat}}$ in a
broad range from 250 to 2500~cm$^{-1}$ (30 -- 300~meV).  This makes
ZrSiS similar to (quasi)two-dimensional Dirac electron systems, such
as graphite and graphene. We assign the flat optical conductivity to
the transitions between quasi-two-dimensional Dirac bands near the
Fermi level. In contrast to graphene, $\sigma_{\rm{flat}}$ is not
supposed to be universal but related to the length of the nodal line
in the reciprocal space, $k_{0}$. When $\sigma_{\rm{flat}}$ and
$k_{0}$ are connected by a simple model, we find good agreement
between experiment and theory. Due to the spin-orbit coupling, the
discussed Dirac bands in ZrSiS possess a small gap $\Delta$, for
which we determine an upper bound max($\Delta$) = 30 meV from our
optical measurements. At low temperatures the momentum-relaxation
rate collapses, and the characteristic length scale of momentum
relaxation is of the order of microns below 50 K.

\end{abstract}

\maketitle

Additionally to the three-dimensional (3D) Dirac and Weyl semimetal
phases, the line-node semimetals (LNSM) proposed in
2011~\cite{Burkov2011} attract more and more
attention~\cite{Kim2015, Ramamurthy2017, Fang2015, Bian2016, Wu2016,
Yu2015, Chen2015, Okamoto2016, Xie2015, Xu2015, Schoop2016,
Neupane2016, Topp2016, CChen2017}. Unlike the point nodes in Dirac
and Weyl semimetals, the linear-band crossings in LNSMs form
continuous lines (loops) in reciprocal space. Most recently, it has
been shown theoretically that in general the topology of nodal lines
within a Brillouin zone (BZ) may be very complex, e.g., nodal lines
may be linked and knotted in different ways~\cite{WChen2017,
Yan2017, Ezawa2017}. The presence of line nodes effectively reduces
the dimensionality of the Dirac bands. Thus, a (3D) LNSM is supposed
to host 2D Dirac electrons. Such 3D materials with 2D Dirac
electrons (i.e. the 3D analogues of graphene) are appealing for
both, basic and applied, research, as they are supposed to
demonstrate a number of unusual electronic properties that can be
useful for potential applications~\cite{Burkov2011, Kim2015,
Ramamurthy2017}. It is worth noting that the LNSM state can arise in
materials with as well as without spin-orbit coupling (SOC)
\cite{Fang2015}.

An evidence for possible realization of a LNSM state has been
recently obtained via angle-resolved photoemission spectroscopy
(ARPES) in PbTaSe$_{2}$~\cite{Bian2016}. A state similar to LNSM,
but with Dirac arcs instead of closed loops, is reported in
PtSn$_{4}$~\cite{Wu2016}. Also, many theoretical propositions are
around for materials realizing the LNSM phase, including
Cu$_{3}$PdN~\cite{Yu2015}, SrIrO$_{3}$~\cite{Chen2015}, CaAgP and
CaAgAs~\cite{Okamoto2016}, as well as a new crystallographic form of
Ca$_{3}$P$_{2}$~\cite{Xie2015}.

Much attention is currently paid to ZrSiS~\cite{Xu2015, Schoop2016,
CChen2017, Neupane2016} and its structural analogues, such as, e.g.,
HfSiS~\cite{CChen2017, Takane2016}, ZrSiTe~\cite{Topp2016, Hu2016},
and GdSbTe~\cite{Hosen2017}. The presence of Dirac bands in ZrSiS
and its family is well established by several experimental methods,
including ARPES~\cite{Schoop2016, Topp2016, Wang2016, Neupane2016,
CChen2017}, Hall measurements~\cite{Singha2016, Sankar2017}, and
quantum oscillations~\cite{Wang2016, Ali2016, Hu2016, Singha2016,
Pezzini2017, Sankar2017, Matusiak2017, Hu2017}, as well as by
electronic-structure calculations~\cite{Schoop2016, Topp2016,
Wang2016, Neupane2016, CChen2017}. These studies demonstrate that
ZrSiS possesses two types of line nodes. The line nodes of the first
type are situated far away ($\sim 0.7$ eV) from the Fermi level; we
dub them as high-energy nodes. Turning on the SOC opens a gap along
certain portions of this nodal line~\cite{Schoop2016, CChen2017}.
The line nodes of the second type appear close to the Fermi level
(low-energy nodes), but are believed to be fully gapped due to SOC,
similarly to such LNSM candidates as Cu$_{3}$PdN~\cite{Yu2015} and
SrIrO$_{3}$~\cite{Chen2015}. The gap, however, is calculated to be
very small, of the order of 10 meV~\cite{Schoop2016}. Such a small
value has indeed been confirmed by recent ARPES
measurements~\cite{CChen2017}, although the resolution was not
sufficient to accurately determine the gap size. At higher energies
(up to a few hundreds meV), the linearity of the line-node Dirac
bands in ZrSiS remains uncompromised~\cite{Schoop2016, CChen2017}.
These low-energy line nodes are in focus of the present study.

Dirac electrons in solids are known to manifest themselves in
peculiar ways in different experiments~\cite{Neto2009, Hasan2010,
Wehling2014, Vafek2014, Armitage2017}. One of such manifestations is
in their optical response (i.e., the ac transport), usually
expressed in terms of the complex optical conductivity,
$\sigma(\omega) = \sigma_{1}(\omega) + i\sigma_{2}(\omega)$. For
example, in the simplest case of electron-hole symmetric
$d$-dimensional ungapped Dirac (Weyl) bands, $\sigma_1(\omega)$ is
supposed to follow a power law, $\sigma_1(\omega) \propto
\omega^{d-2}$~\cite{Hosur2012, Bacsi2013}.

This optical-conductivity behavior --~unusual for conventional
materials~-- has indeed been confirmed for quasi-2D electrons in
graphite and graphene, where $\sigma_1(\omega) \approx
\rm{const}(\omega)$ \cite{Kuzmenko2008, Mak2008}, and for 3D Dirac
electrons in such point-node Dirac/Weyl semimetals as ZrTe$_{5}$
\cite{Chen2015ZrTe5}, Cd$_{3}$As$_{2}$ \cite{Neubauer2016}, and TaAs
\cite{Xu2016}, where $\sigma_1(\omega) \propto \omega$ was reported.
As mentioned above, the Dirac electrons in a (3D) LNSM live
effectively in two dimensions. Thus, the optical conductivity of a
LNSM should be similar to the one of graphene, i.e.
frequency-independent. Such flat optical conductivity in LNSMs has
indeed been predicted by theory recently~\cite{Carbotte2017,
Mukherjee2017, Ahn2017}.

Here, we report observation of frequency-independent optical
conductivity in ZrSiS. This evidences the existence of quasi-2D
Dirac states and a quasi-2D electronic ac transport in this
material. From our optical measurements, we extract the length of
the nodal line (the node is understood here as the Dirac point of
the gapped Dirac band) and estimate the size of the gap in this
band.

The investigated single crystals were grown by loading equimolar
amounts of Zr, Si, and S together with a small amount of iodine in a
sealed quartz tube, which was kept at 1100 $^{\circ}$C for 1 week. A
temperature gradient of 100 $^{\circ}$C was applied and the crystals
were collected at the cold end of the tube. The crystal structure
(tetragonal, space group \textit{P}4/\textit{mmm}) was confirmed
with x-ray and electron diffraction similarly to
Ref.~\cite{Schoop2016}.

\begin{figure}[b]
\centering
\includegraphics[width=\columnwidth]{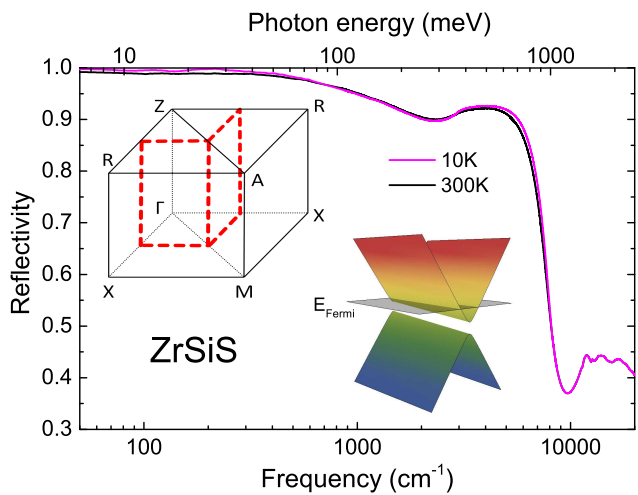}
\caption{Frequency-dependent reflectivity of ZrSiS at 10 and 300 K.
The reflectivity curves measured at intermediate temperatures lie in
between the two curves and are not shown for clarity. No temperature
dependence is seen above $\sim 8000$ cm$^{-1}$. The sketches show
the position of the low-energy nodal line (red dashed line) in BZ
and the Dirac bands near the Fermi level.} \label{R}
\end{figure}

The optical reflectivity $R(\nu)$ was measured at 10 to 300 K over a
broad frequency range from $\nu = \omega/2\pi \approx 50$ to 25000
cm$^{-1}$ using commercial Fourier-transform infrared spectrometers.
All measurements were performed on freshly cleaved (001) surfaces.
In accordance with the tetragonal structure, no in-plane optical
anisotropy was detected. At low frequencies, an in-situ gold
evaporation technique was utilized for reference measurements, while
above 1000 cm$^{-1}$ gold and protected silver mirrors served as
references. The high-frequency range was extended by
room-temperature ellipsometry measurements up to 45 000 cm$^{-1}$ in
order to obtain more accurate results for the Kramers-Kronig
analysis~\cite{Dressel2002}. The Kramers-Kronig analysis was made
involving the x-ray atomic scattering functions for high-frequency
extrapolations~\cite{Tanner2015} and dc-conductivity values,
$\sigma_{dc} (T)$, and the reflectivity-fitting
procedure~\cite{Schilling2017} for zero-frequency extrapolations.
Important to note, that our optical measurements reflect the bulk
material properties, since the penetration depth is above 40 nm for
any measurement frequency.

\begin{figure}[t]
\centering
\includegraphics[width=\columnwidth]{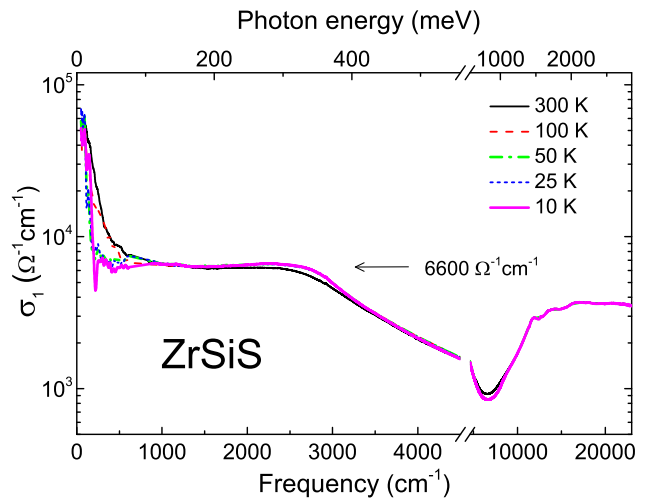}
\caption{Real part of the optical conductivity of ZrSiS as a
function of frequency.} \label{sg1}
\end{figure}

\begin{figure}[t]
\centering
\includegraphics[width=0.9\columnwidth]{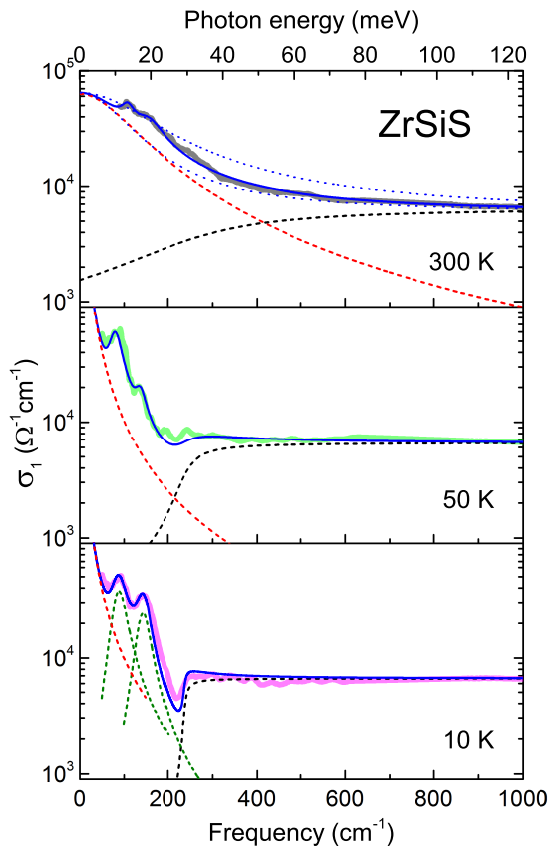}
\caption{Examples of the optical conductivity fits for ZrSiS at low
frequencies. Thick solid lines are experimental data; thin solid
lines are total fits; dashed lines represent fit contributions of
the Drude (red), Pauli-edge (black), and Lorentzian (olive, shown
only for T =10 K) terms. Thin dotted (blue) lines represents
attempts to fit the 300-K data without the Lorentzians.}
\label{fits}
\end{figure}

\begin{figure}[]
\centering
\includegraphics[width=\columnwidth]{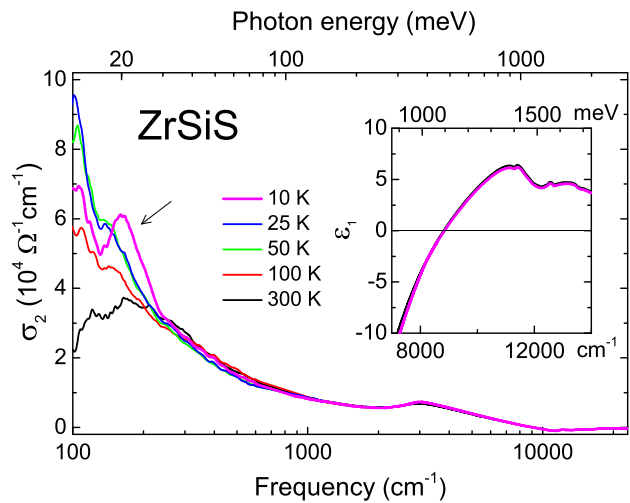}
\caption{Main frame: imaginary part of the optical conductivity in
ZrSiS. Arrow indicates the position of the Pauli edge at 10 K.
Inset: real part of the dielectric constant near the plasma
frequency.} \label{sg2}
\end{figure}

The measured frequency-dependent reflectivity $R(\nu)$ is shown in
Fig.~\ref{R} for selected temperatures. Above 1000 cm$^{-1}$, the
temperature has only a minor influence on the spectra. In the
low-frequency range, the reflectivity is rather high (above $99\%
$), in agreement with the very low dc resistivity~\cite{Singha2016,
Ali2016}. The results of the Kramers-Kronig analysis are shown in
Figs.~\ref{sg1} -- \ref{sg2} in terms of the real and imaginary
parts of optical conductivity, as well as the real part of
permittivity, $\varepsilon_{1}(\omega) = 1 - 4\pi \sigma_{2}(\omega)
/ \omega$.

An important result of this work is presented in Fig.~\ref{sg1}: the
real part of optical conductivity is almost frequency-independent,
$\sigma_{1}(\omega) = \sigma_{\rm{flat}} \approx$ 6600
$\Omega^{-1}$cm$^{-1}$, in the range from 250 to 2500 cm$^{-1}$ [30
-- 300 meV] basically at all temperatures investigated (at $T \geq
100$ K, the flat region starts at a bit higher frequencies because
of a rather broad free-electron contribution). Such
frequency-independent behavior of $\sigma_{1}(\omega)$ is similar to
what has been predicted~\cite{Ando2002} and observed~\cite{Mak2008}
in graphene and matches the theory for the optical response due to
transitions between the 2D Dirac states in LNSMs~\cite{Carbotte2017,
Ahn2017, Mukherjee2017}. In contrast to graphene, in LNSMs no
universal sheet conductance is expected and $\sigma_{\rm{flat}}$ is
related instead to the length of the nodal line $k_{0}$ in a BZ. For
a circular nodal line, one has:
\begin{equation}
\sigma_{1}(\omega) = \sigma_{\rm{flat}} = \frac{e^2 k_{0}} {16
\hbar}. \label{k_0}
\end{equation}
It is assumed here that the plane of the nodal circle is
perpendicular to the electric-field component of the probing
radiation and that there is no particle-hole
asymmetry~\cite{Carbotte2017, Ahn2017, Mukherjee2017}. In ZrSiS, the
low-energy nodal line is not circular and not even flat; instead the
BZ contains a 3D ``cage'' of nodal lines~\cite{Schoop2016}. Thus, a
straightforward application of this formula is not rigourously
validated. Nevertheless, having no better model at hand, we use
Eq.~(\ref{k_0}) for a rough estimate of $k_{0} = 4.3$ {\AA}$^{-1}$.
This value seems to be reasonable: according to the band-structure
calculations, the total lengths of the nodal line projections on the
[100] and [001] directions are about 3.5 and 6 {\AA}$^{-1}$ per BZ,
respectively.

At $\nu > 3000$ cm$^{-1}$, $\sigma_1(\omega)$ is not
frequency-independent anymore; it decreases with frequency. This is
likely because the 2D Dirac band looses its linearity at such large
energies~\cite{Schoop2016}. At even higher frequencies,
$\sigma_1(\omega)$ starts to rise as $\omega$ increases. It would be
of interest to detect the transitions between the high-energy Dirac
bands above approximately 1.3~eV = 10\,500~cm$^{-1}$, but we do not
see any clear signatures of such transitions: as discussed above,
these Dirac bands are partly gapped by SOC and for bands of such
complex shapes the optical conductivity is not expected to be flat
or linear in frequency. Additionally, non-Dirac bands above the
Fermi level also contribute to the absorption processes at these
frequencies, as one can see from the band structure calculations of
Refs.~\cite{Schoop2016, CChen2017}.

At the lowest frequencies measured, $\nu < 500$~cm$^{-1}$,
$\sigma_1(\omega)$ also deviates from the flat behavior, as can be
seen in Figs.~\ref{sg1} and \ref{fits}, and exhibit several
features. The contribution of free carriers (so-called Drude
response) is present at all temperatures, but best seen at 100 and
300 K. A free-electron component is expected in the optical response
because the Fermi level in ZrSiS is slightly above the nodal
line~\cite{Schoop2016, CChen2017}. As the temperature drops, the
Drude band narrows, revealing two distinct modes at around 100 and
150~cm$^{-1}$, which may be explained by some sort of electron
localization; an elaborate discussion of the modes is outside the
scope of this paper. The narrowing of the Drude band reflects the
strong suppression of the dc resistivity~\cite{Singha2016, Ali2016}
and of the carrier scattering rate (or, more accurately, the
momentum-relaxation rate) as $T \rightarrow 0$.

At the lowest temperatures, $\sigma_1(\omega)$ develops a minimum
around 200~cm$^{-1}$ (bottom panel of Fig.~\ref{fits}). We relate
this dip and the corresponding feature in $\sigma_2(\omega)$, see
Fig.~\ref{sg2}, to the Pauli blocking of the transitions in the 2D
band. Such features, related to the position of the Fermi level, are
well known in semiconductors~\cite{Yu2010} and have recently been
discussed in relation to graphene and Dirac/Weyl
semimetals~\cite{Gusynin2007, Kotov2016, Jenkins2016}. As already
mentioned, band-structure calculations and ARPES measurements locate
the Fermi level in ZrSiS in the upper (conduction) Dirac
band~\cite{Schoop2016, CChen2017}, see the sketch in Fig.~\ref{R}.
Thus, the Pauli edge must be seen in the interband portion of
optical conductivity. An onset of the interband transitions commonly
shows up at the frequency equal to $\max\{\Delta,
2\mu\}$~\cite{Ahn2017, Mukherjee2017, Neubauer2016, Gusynin2007,
Kotov2016, Jenkins2016}, with $\Delta$ being the band gap and $\mu$
the position of the Fermi level relative to the Dirac point. Thus,
Eq.~(\ref{k_0}) can be modified to:
\begin{equation}
\sigma_{1}(\omega) = \frac{e^2 k_{0}} {16 \hbar} \times \theta
(\hbar\omega - \max\{\Delta, 2\mu\}), \label{onset}
\end{equation}
where $\theta (x)$ is the Heaviside step function. From
Eq.~(\ref{onset}) and the bottom panel of Fig.~\ref{fits}, one can
conclude that $\max\{\Delta, 2\mu\}$ must be smaller than
approximately 250~cm$^{-1}$ or some 30~meV and, hence $\Delta <
30$~meV.

This estimate of the upper limit of $\Delta$ from optical data is in
good agreement with the value obtained from band-structure
calculations (15 meV, \cite{Schoop2016}). The ARPES value of 60 meV
\cite{CChen2017} likely overestimates the gap due to relatively low
ARPES resolution. The fact that the relative position of the Dirac
points and the Fermi level is slightly
$\textbf{k}$-dependent~\cite{Schoop2016, CChen2017} might lead to a
broadening of the optical Pauli-edge feature observed. Importantly,
even if the Fermi level appears within the gap for some values of
$\textbf{k}$, our conclusion on the upper limit of $\Delta$ still
holds.

To get some more quantitative estimates of the parameters
determining the optical response, we fit the optical conductivity
with a model consisting of a Drude term, two Lorentzians, and a term
describing the Pauli edge. Scattering and other processes, leading
to broadening of the sharp step in Eq.~(\ref{onset}), may be taken
into account by replacing the Heaviside function with
\begin{equation}
\frac{1}{2}+\frac{1}{\pi}\arctan \frac{\omega - \max\{\Delta,
2\mu\}/\hbar}{\Gamma}, \label{arctangent}
\end{equation}
for example, where $\Gamma$ represents a broadening parameter due to
$\textbf{k}$-dependent gap, impurity scattering, or temperature. At
$T=10$~K, reasonable fits can be obtained with a very sharp Pauli
edge, i.e.\ with $\Gamma$ of a few cm$^{-1}$. We set
$\Gamma=k_{B}T/\hbar$ for all temperatures, since smaller values
seem not to be physical. This yields $\Gamma = 7$, 35, and
210~cm$^{-1}$ for 10, 50, and 300 K, respectively, see
Fig.~\ref{fits}.

In all our fits we keep the zero-frequency limit of the Drude term
equal to $\sigma_{dc}$ at all temperatures. Owed to the broad Drude
tail, the description of the 300-K data is straightforward. It
provides the momentum-relaxation rate of free carriers, $\gamma =
1/(2\pi \tau) = (120 \pm 10)$ cm$^{-1}$ ($\tau$ is the corresponding
relaxation time), and a plasma frequency, $\omega_{pl}/2\pi = (24000
\pm 1000)$ cm$^{-1}$~\cite{Lorentzian}. On the other hand, the
screened plasma frequency, $\omega_{pl}^{scr} = \omega_{pl} /
\sqrt{\varepsilon_{\infty}}$ ($\varepsilon_{\infty}$ is the
contribution of the higher-frequency optical transitions to
$\varepsilon_{1}$), can be directly determined from optical
measurements as the zero-crossing point of
$\varepsilon_{1}(\nu)$~\cite{Dressel2002}. We find
$\omega_{pl}^{scr} $ to be temperature independent and situated at
8900 cm$^{-1}$, cf.\ the inset of Fig.~\ref{sg2}. Hence,
$\varepsilon_{\infty} = (\omega_{pl} / \omega_{pl}^{scr})^{2}
\approx 7$, which is in good agrement with the optical measurements,
presented in same figure.

As one can see from Figs.~\ref{sg1} -- \ref{sg2}, the Drude term
becomes narrower as $T \rightarrow 0$. At low temperatures, $\gamma$
is below our measurement window and our fits thus might become
ambiguous. To avoid this, we first tried to keep the plasma
frequency of the Drude term constant as a function of $T$; but this
turned out to be unsatisfactory. Some spectral weight had to be
redistributed between the Drude term and the Lorentzians.
Nevertheless, we tried to have this spectral weight transfer as
small as possible and the total plasma frequency of the three terms
(Drude plus two Lorentzians) to be temperature-independent in
accordance with the temperature-independent $\omega_{pl}$. Examples
of the fits obtained in this way are shown in Fig.\ref{fits}. At $T
\leq 50$ K, we found that $\gamma\approx 2$ to 2.5~cm$^{-1}$ and the
momentum-relaxing $\tau$ is 2.1 to 2.7 ps.

Interestingly, at low temperatures the momentum-relaxation length,
$\ell_{\rm{mr}}=v_{F}\tau$, obtained from our estimate of $\tau$,
becomes macroscopically large. Using $v_{F} = 5\times10^{5}$~m/s as
an average Fermi velocity in the low-energy Dirac
bands~\cite{Singha2016}, we obtain $\ell_{\rm{mr}} \geq 1$ $\mu$m
for $T \leq 50$ K. This implies that the hydrodynamic behavior of
electrons, reported recently in clean samples of
graphene~\cite{Bandurin2016, Crossno2016} and the Weyl semimetal
WP$_{2}$~\cite{Gooth2017}, might also be realized in ZrSiS. This
proposition seems reasonable, because only linear bands with highly
mobile carriers (typical mobilities are $10^3$ to $10^4$ cm$^{2}$/Vs
\cite{Singha2016, Sankar2017, Hu2017}) cross the Fermi level in
ZrSiS.

Summarizing, the real part of the optical conductivity of ZrSiS was
found to be independent of frequency in a rather broad range from
250 to 2500 cm$^{-1}$ (30 -- 300 meV). Our observations are
supported by recent theoretical predictions for the optical response
of LNSMs, and constitute and independent confirmation of 2D Dirac
bands in ZrSiS near the Fermi level. The characteristic features of
the Pauli edge, appearing in the low-frequency spectra, provide the
upper limit (250 cm$^{-1}$, 30 meV) for the gap between the 2D Dirac
bands. The momentum-relaxation length is at the micrometer scale at
$T \leq 50$ K. Overall, our optical measurements reveal that ZrSiS
is a gapped line-node semimetal with the electronic properties
determined primarily by 2D Dirac electrons with rather slow momentum
relaxation at low temperatures.

We thank Jules P. Carbotte, David Neubauer, and Raquel Queiroz for
fruitful discussions and Gabriele Untereiner and Ievgen Voloshenko
for technical support. This work was funded by the Deutsche
Forschungsgesellschaft (DFG) via grant No. DR228/51-1. L. M. S. was
supported by Minerva fast track scholarship from Max Planck Society.


\begin{thebibliography}{99}
\bibitem{Lorentzian} Note, even at $T=300$~K, the Lorentzians
in our model improve the fit, see Fig.~\ref{fits}(a).
\end{thebibliography}
\end{document}